\documentclass{article}
\usepackage[T1]{fontenc}
\usepackage{iclr2027_conference,times}
\iclrpreprint

\usepackage{amsmath,amsfonts,bm}

\def\eqref#1{equation~\ref{#1}}

\def\1{\bm{1}}

\DeclareMathAlphabet{\mathsfit}{\encodingdefault}{\sfdefault}{m}{sl}
\SetMathAlphabet{\mathsfit}{bold}{\encodingdefault}{\sfdefault}{bx}{n}

\newcommand{\E}{\mathbb{E}}

\newcommand{\R}{\mathbb{R}}

\usepackage{graphicx}
\usepackage{booktabs}
\usepackage{multirow}
\usepackage{array}
\usepackage{xcolor}
\usepackage{colortbl}
\usepackage{amsmath,amssymb}
\usepackage{hyperref}
\hypersetup{hidelinks}
\usepackage{url}
\usepackage{enumitem}
\usepackage{xspace}
\newcommand{\PtwoP}{\textsc{P2P}\xspace}

\newcommand{\loss}{\mathcal{L}}

\definecolor{lightblue}{RGB}{235,243,250}
\definecolor{lightgray}{RGB}{245,245,245}

\title{P2P: Cross-View Population Denoising for Unpaired Single-Cell Perturbation Response Prediction}

\author{Haojie Yang \\
College of Intelligence and Computing \\
Tianjin University, Tianjin, China \\
\texttt{haojae@tju.edu.cn}
\And
Ran Su$^*$ \\
College of Intelligence and Computing \\
Tianjin University, Tianjin, China \\
\texttt{ran.su@tju.edu.cn} \\
$^*$Corresponding author}

\begin{document}
\maketitle

\begin{figure}[h]
\centering
\includegraphics[width=\linewidth]{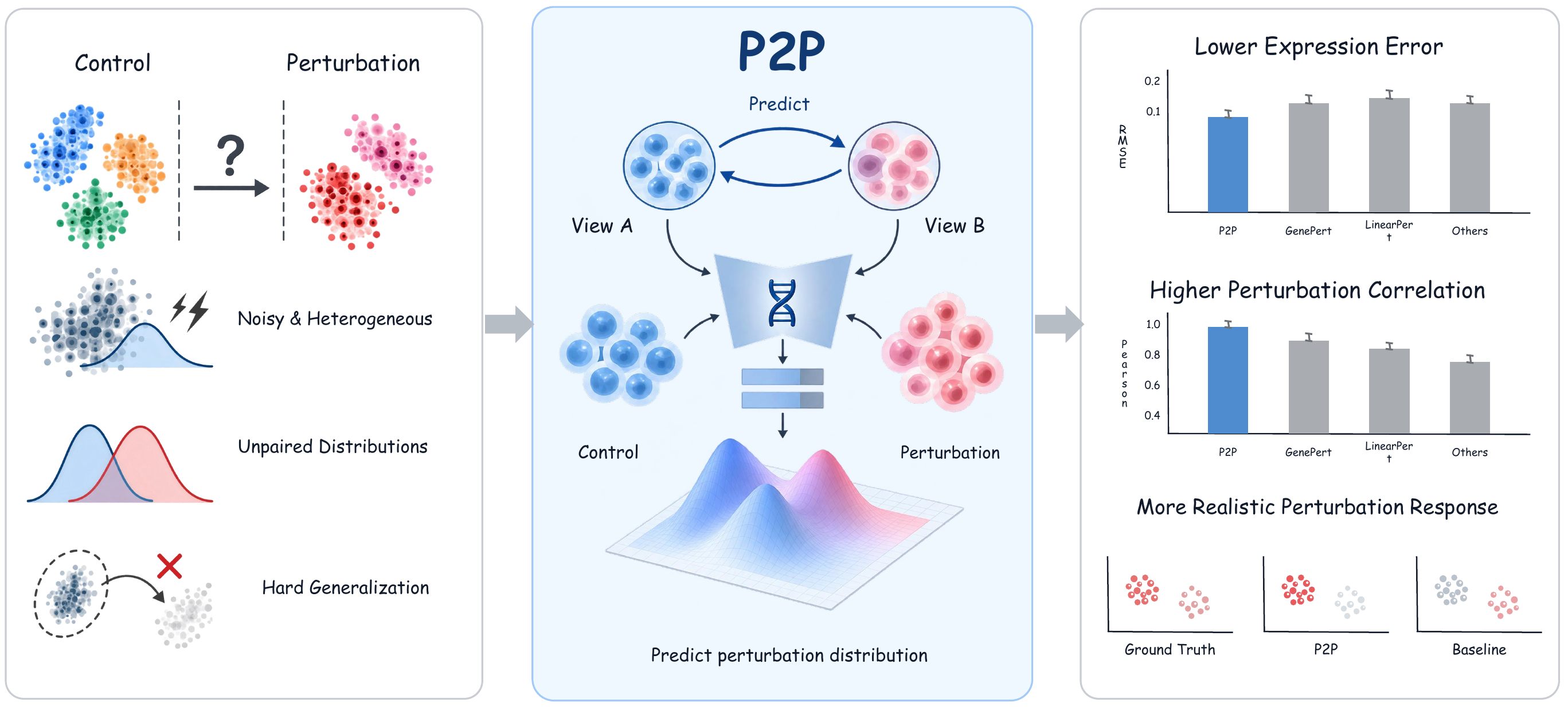}
\caption{Cross-view population denoising. Perturbation assays provide noisy and heterogeneous cell collections without one-to-one control pairs. \PtwoP samples two stochastic views from each population and learns the shared condition-level effect, which is then evaluated through population expression, effect recovery, and response structure.}
\label{fig:motivation}
\end{figure}

\begin{abstract}
AIVC (AI Virtual Cell) is a learned simulator of cellular behavior across conditions. Predicting how a cell population responds transcriptionally to a genetic perturbation is a core task. Perturb-seq records that response by destructive sequencing, so a control cell and a perturbed cell are never observed as a pair, and cells under one condition remain heterogeneous and noisy. Regression on individual cells absorbs sampling variation into the estimated effect, whereas interpretation requires the reproducible population effect. \PtwoP (Perturbation-to-Perturbation) takes a stochastic cell-set view as its supervision unit. Two views drawn from the same condition share a reproducible population effect and differ by view-specific variation. A permutation-invariant set encoder summarizes the control population, a structured encoder represents perturbation tokens, cellular context, dose, and combination interactions, and a gate blends empirical condition-effect memory with a neural residual. A heteroscedastic head predicts the population mean and gene-wise response variance. Under one protocol and five seeds, \PtwoP attains the lowest expression RMSE and the highest Effect Pearson, DEG F1, and DEG average precision on each of Adamson, Norman, Replogle K562, and Replogle RPE1 relative to GenePert, LinearPert, SLIM, Scouter, and scPILOT. On Replogle K562, Effect Pearson rises from 0.643 to 0.702 and DEG F1 rises from 0.067 to 0.178 relative to Scouter, the strongest baseline on both metrics.
\end{abstract}

\section{Introduction}
Single-cell RNA sequencing measures expression variation across thousands of cells \citep{macosko2015dropseq,zheng2017digital}. Coupling these readouts to pooled genetic screens connects interventions to cellular responses \citep{dixit2016perturbseq,adamson2016multiplexed,datlinger2017cropseq,hastie2021fluidicindex}. Combinatorial and genome-scale screens extend this approach to genetic interactions and broad genotype-phenotype landscapes \citep{norman2019exploring,replogle2022mapping}. A predictive model can estimate responses for unmeasured conditions and prioritize affected genes for follow-up experiments. The practical target is a population response, since biological conclusions usually rely on many cells.

The data structure complicates this target. RNA sequencing is destructive, so control cells and their perturbed counterparts are not observed before and after treatment. Multiplexed measurements and integration methods address sample identity and technical variation \citep{stoeckius2018cellhashing,haghverdi2018mnn,korsunsky2019harmony}, but cells under one condition still differ in cycle, stress state, and capture efficiency. Pairing an arbitrary control cell with an unrelated perturbed cell creates an unobserved counterfactual. The resulting target mixes the condition effect with sampling variation.

Figure~\ref{fig:motivation} presents the supervision principle behind \PtwoP. Two random sets sampled from one condition have different sampling variation but share a reproducible population response. One set can therefore supervise the population statistics predicted from the other. \PtwoP constructs paired stochastic views of the control and perturbed populations, encodes each unordered cell set through shared functions and symmetric moments, and learns the condition effect through cross-view agreement.

The model combines empirical effect memory with a neural residual for context, dose, and combination structure. Its output contains a population mean and gene-wise variance, separating effect recovery from response heterogeneity. The central contribution is a population learning objective that uses stochastic cell sets as supervision units without requiring cell correspondences. The four-dataset evaluation tests this formulation through expression error, effect correlation, and DEG ranking, while component ablations isolate the roles of cross-view supervision, memory, and consistency constraints.

\section{Related Work}
Perturbation models reconstruct different aspects of an intervention. Statistical models link pooled CRISPR assignments to transcriptional changes \citep{weber2019crispromodel}. scGen and CPA learn latent responses to perturbations and covariates \citep{lotfollahi2019scgen,lotfollahi2020cpa}, while GEARS incorporates gene relationships to predict multigene responses \citep{roohani2023gears}. These methods motivate structured representations of the intervention. \PtwoP addresses how the response target is constructed from finite cell samples and uses agreement between sampled populations as a learning signal.

Probabilistic representations capture variation beyond a population mean. scVI and scvi-tools provide latent-variable models and inference infrastructure for single-cell data \citep{lopez2018scvi,gayoso2022scvitools}. totalVI and MultiVI extend this framework to multimodal observations \citep{gayoso2021totalvi,ashu2023multivi}, while scGPT learns gene representations through generative pretraining \citep{cui2024scgpt}. \PtwoP couples a mean and variance readout to an effect-centered objective, using repeated population views to supervise the reproducible response.

Integration and reference mapping establish comparable cell representations across datasets. Seurat and scArches align datasets or map queries to reference atlases \citep{stuart2019integration,lotfollahi2022scarches}; Cell BLAST supports reference search through cell embeddings \citep{cao2020cellblast}. Scanpy and UMAP provide analysis and visualization tools \citep{wolf2018scanpy,mcinnes2018umap}. Integration benchmarks evaluate both technical alignment and conservation of biological structure \citep{luecken2022scib}. Population perturbation prediction adds a distinct requirement: the representation must preserve the signed response relative to a matched control population.

Biological structure can also be represented through regulatory networks, propagation, and cell-state dynamics. SCENIC infers regulatory programs, and SCAVENGE propagates variant-associated signals across cell networks \citep{aibar2017scenic,yu2022scavenge}. scVelo and CellRank use dynamical information to characterize state transitions and fate probabilities \citep{bergen2020scvelo,lange2022cellrank}. Graph neural networks encode relational structure \citep{wu2021gnnsurvey}, while symmetric aggregation supports learning from unordered sets \citep{qi2017pointnet}. \PtwoP uses the latter principle to encode cell populations and evaluates recovered effects through population, gene, and graph-derived views.

\section{Method}
\subsection{Problem formulation and cross-view supervision}
Let $G$ be the number of measured genes. For a context $s$, dose $d$, and perturbation token set $p$, let $X_{0,s}$ denote a control cell set and $X_{p,s,d}$ the corresponding perturbed set, with each cell in $\R^G$. The population mean and the condition effect are defined in Eq.~\ref{eq:population-effect}.
\begin{equation}
\mu_{p,s,d}=\E[x\mid p,s,d],\qquad
\delta_{p,s,d}=\mu_{p,s,d}-\mu_{0,s}.
\label{eq:population-effect}
\end{equation}
The observations contain no natural mapping between elements of $X_{0,s}$ and $X_{p,s,d}$. Two stochastic views are therefore drawn for each set using Eq.~\ref{eq:view-split},
\begin{equation}
(X_{0,s}^{a},X_{0,s}^{b},X_{p,s,d}^{a},X_{p,s,d}^{b})\sim \operatorname{ViewSplit}(X_{0,s},X_{p,s,d}).
\label{eq:view-split}
\end{equation}
When enough cells are available, the views are disjoint samples. For smaller conditions, the implementation uses independent sampling with replacement so that the prescribed view size remains available.

A training prediction maps one control view and condition metadata to the mean of the other perturbed view as specified in Eq.~\ref{eq:cross-view-prediction},
\begin{equation}
\widehat{\mu}^{a\leftarrow b}=f_\theta(X_{0,s}^{b},p,s,d),
\label{eq:cross-view-prediction}
\end{equation}
with the reverse direction computed in the same update. The loss compares the prediction with the target view mean, its effect relative to the matching control view, and the within-view second moment. At inference, the target view is absent. The model receives a control set, perturbation tokens, context, dose, and an optional condition memory calculated from training-role cells.

This formulation distinguishes three quantities that are often conflated. A cell-level state is one observed expression vector. A population mean is a statistic of a condition. The perturbation effect is a difference between two population means under a common context. \PtwoP optimizes the latter two quantities and uses cell-level observations only to construct the set statistics.

\subsection{P2P architecture}
\begin{figure}[t]
\centering
\includegraphics[width=0.98\linewidth]{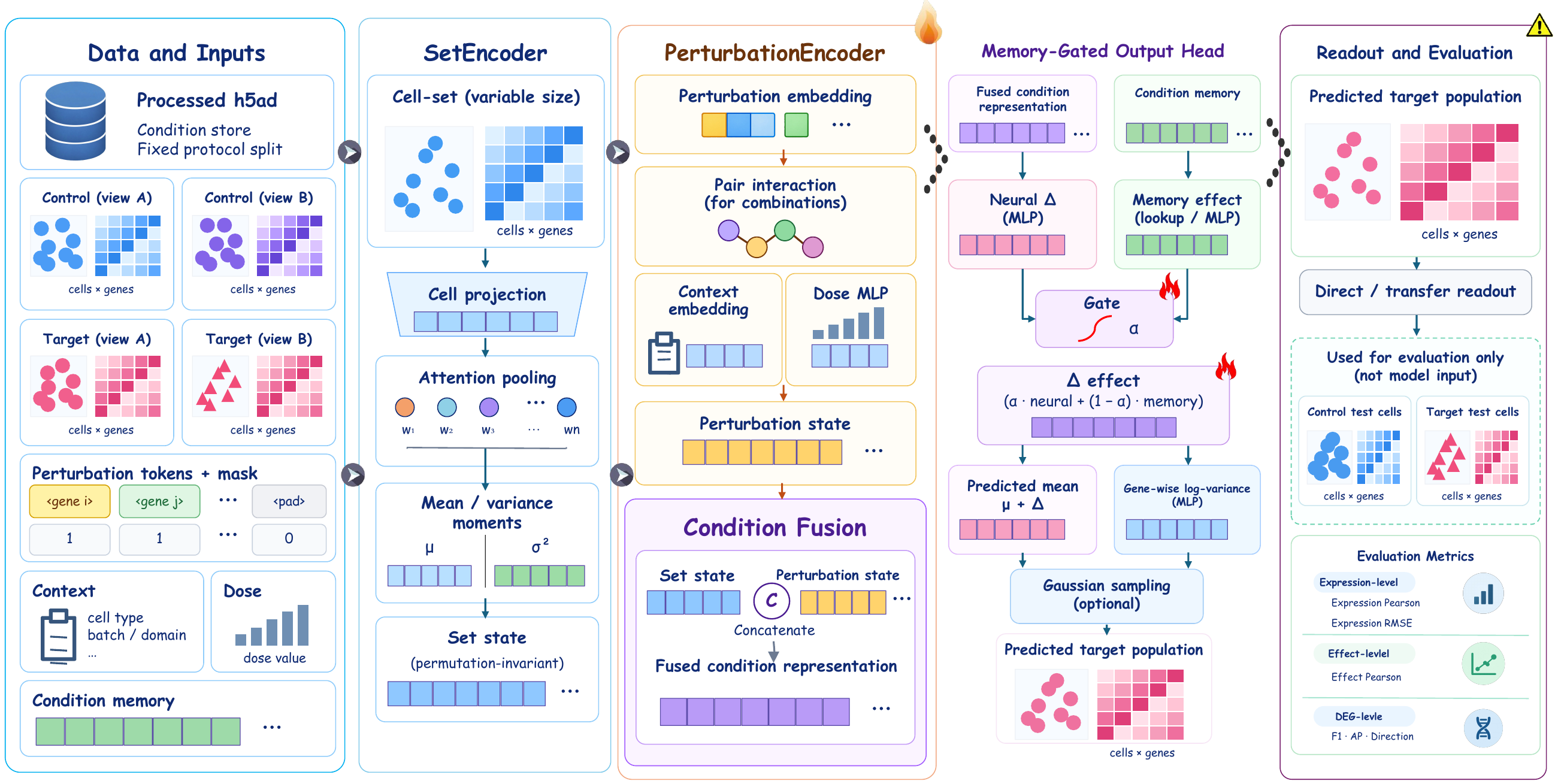}
\caption{P2P architecture and readout. The permutation-invariant SetEncoder summarizes a control cell set, while the structured perturbation encoder combines perturbation tokens, context, dose, and pair interactions. A memory-gated effect head combines the empirical condition effect with a neural residual and predicts the population mean and gene-wise log variance. Target views supervise the cross-view objective during training.}
\label{fig:architecture}
\end{figure}

The complete data path is shown in Figure~\ref{fig:architecture}; the subsections below specify the components that determine its population effect prediction.

\subsubsection{Permutation-invariant population encoding}
For a cell set $X=\{x_i\}_{i=1}^{n}$, a shared projection maps each cell to a hidden state according to Eq.~\ref{eq:cell-projection},
\begin{equation}
h_i=\operatorname{Dropout}\!\left[\operatorname{GELU}\!\left(\operatorname{LN}(W_xx_i+b_x)\right)\right].
\label{eq:cell-projection}
\end{equation}
A score network produces normalized cell weights $a_i$. The set state includes the attention-weighted summary and the unweighted dispersion summary in Eq.~\ref{eq:set-moments},
\begin{equation}
\bar h_a=\sum_i a_i h_i,\qquad
\bar h=\frac{1}{n}\sum_i h_i,\qquad
v=\frac{1}{n}\sum_i(h_i-\bar h)^2,
\label{eq:set-moments}
\end{equation}
\begin{equation}
h_{set}=\operatorname{MomentMLP}([\bar h_a,v]).
\label{eq:set-state}
\end{equation}
The final set state is given by Eq.~\ref{eq:set-state}. Every operation is shared across cells and followed by a symmetric sum, mean, or variance. The resulting encoder is invariant to permutations of the input cells, while $v$ preserves a compact representation of population heterogeneity. The unweighted mean anchors the measured control baseline, while attention captures informative heterogeneity.

\subsubsection{Structured perturbation and context encoding}
A perturbation can contain one token or several tokens. The token embeddings are pooled with a mask as in Eq.~\ref{eq:perturbation-pool}, so padding cannot contribute to the condition state,
\begin{equation}
e_p=\frac{\sum_j m_jE[p_j]}{\max(1,\sum_jm_j)}.
\label{eq:perturbation-pool}
\end{equation}
For a combination, a small pair-interaction MLP adds a nonlinear correction $q(e_p)$; the correction is disabled for a single token. Context and dose are encoded separately, and the condition state in Eq.~\ref{eq:condition-state} is
\begin{equation}
h_{pert}=\operatorname{LN}\left(e_p+\mathbb{1}_{|p|>1}q(e_p)+E_s[s]+r(d)\right).
\label{eq:condition-state}
\end{equation}
This design exposes the factors that change a population response and supports both single-gene and combination conditions, while keeping context and dose available to the residual branch.

\subsubsection{Memory-gated effect prediction}
The control set state and structured condition state are fused by a two-layer MLP, producing the residual in Eq.~\ref{eq:neural-residual},
\begin{equation}
h_{cond}=\operatorname{Fusion}(h_{set},h_{pert}),\qquad
\Delta_{neural}=\operatorname{DeltaHead}(h_{cond}).
\label{eq:neural-residual}
\end{equation}
For a condition with training-role cells, an empirical memory is computed as the population difference in Eq.~\ref{eq:condition-memory},
\begin{equation}
m_{p,s,d}=\operatorname{mean}(X_{p,s,d}^{role})-\operatorname{mean}(X_{0,s}^{role}).
\label{eq:condition-memory}
\end{equation}
A learned scalar gate combines this stable statistic with the neural residual according to Eq.~\ref{eq:gated-effect},
\begin{equation}
g=\sigma(\alpha),\qquad
\widehat{\delta}=g\,m_{p,s,d}+(1-g)\,\Delta_{neural},qquad
\widehat{\mu}=\operatorname{mean}(X_{0,s})+\widehat{\delta}.
\label{eq:gated-effect}
\end{equation}
The gate is initialized with $\alpha=1.1$, which gives the memory branch an initial weight of approximately $0.75$ while retaining a trainable residual. The formulation makes the statistical prior explicit. The explicit prior lets the model interpolate between empirical condition effects and the learned residual. Transfer readouts use target training cells to construct the empirical memory and are reported under the memory-assisted protocol.

\subsubsection{Heteroscedastic output}
A second head predicts a gene-wise log variance using Eq.~\ref{eq:log-variance},
\begin{equation}
\log\widehat{\sigma}^{2}=\operatorname{clip}\left(\operatorname{LogVarHead}(h_{cond}),-8,4\right).
\label{eq:log-variance}
\end{equation}
Optional population samples are drawn as $\widetilde{Y}=\widehat{\mu}+\epsilon\exp(\frac{1}{2}\log\widehat{\sigma}^{2})$ with standard normal $\epsilon$. The sample mean is centered on $\widehat{\mu}$ before evaluation. This head is trained jointly with the mean and moment losses and is used to represent response variation; uncertainty quality is evaluated separately from point prediction.

\subsubsection{Effect-centric cross-view objective}
For a target view with mean $\bar y$ and matching control mean $\bar c$, define the predicted effect as $\widehat\delta=\widehat\mu-\bar c$. The direct Gaussian term is defined in Eq.~\ref{eq:nll},
\begin{equation}
\loss_{nll}=\frac{1}{2G}\sum_{g=1}^{G}\left[\log\widehat\sigma_g^2+(\bar y_g-\widehat\mu_g)^2\exp(-\log\widehat\sigma_g^2)\right].
\label{eq:nll}
\end{equation}
The effect term is a weighted Smooth L1 distance between $\widehat\delta$ and $\delta=\bar y-\bar c$, where the gene weight is $1+|\delta_g|/[\operatorname{mean}_k|\delta_k|+\varepsilon]$. An effect-correlation term compares centered effect vectors. A moment term matches the predicted variance to the target-view variance in log space. A consistency term couples the two view predictions for the same condition.

The full objective combines these terms as shown in Eq.~\ref{eq:full-objective},
\begin{equation}
\loss=0.35\loss_{nll}+1.50\loss_{\Delta}+0.15\loss_{corr}+0.08\loss_{mom}+0.05\loss_{cv}.
\label{eq:full-objective}
\end{equation}
For each update, direct pairs and the crossed pairs $X_{0}^{a}\rightarrow X_{p}^{b}$ and $X_{0}^{b}\rightarrow X_{p}^{a}$ are evaluated. Crossed NLL and effect losses enter their direct counterparts with weight $0.25$. The objective therefore gives the model two target views and two control views without treating a cell in one view as the counterpart of a cell in the other.

\subsection{Population interpretation}
The cross-view objective has a simple population interpretation. Suppose that, for a fixed condition, a view mean can be written as $\bar y^{v}=\mu_{p,s,d}+\eta^{v}$ and a control view mean as $\bar c^{v}=\mu_{0,s}+\xi^{v}$. If the view errors have conditional mean zero and the two draws use the same sampling distribution, then the shared first moments satisfy Eq.~\ref{eq:cross-view-unbiased},
\begin{equation}
\E[\bar y^{v}\mid p,s,d]=\mu_{p,s,d},\qquad
\E[\bar y^{v}-\bar c^{v}\mid p,s,d]=\delta_{p,s,d}.
\label{eq:cross-view-unbiased}
\end{equation}
These equalities identify the population first moment targeted by cross-view supervision. When the input is independent of the target-view sampling error, the conditional target has this first moment. The weighted Smooth L1 and correlation terms then shape optimization toward effect magnitude and rank agreement.

The implementation follows this interpretation with a fixed role split and a fixed task list. Training-role cells are used for optimization and memory construction, validation-role cells select checkpoints, and support and test roles are reserved for readout and scoring. For small conditions, view overlap is possible because of replacement sampling; the protocol records overlap explicitly.

\section{Experiments}
\subsection{Datasets and protocol}
The benchmark uses the processed AnnData files supplied with the project. Adamson is a CRISPRi screen focused on the unfolded protein response. Norman contains single and paired CRISPRa perturbations. Replogle K562 and RPE1 are essential-gene CRISPRi screens. The file-level metadata are retained throughout. The Norman processed file carries an A549 cell-type label, and all Norman analyses use that label.

\begin{figure}[t]
\centering
\includegraphics[width=\linewidth]{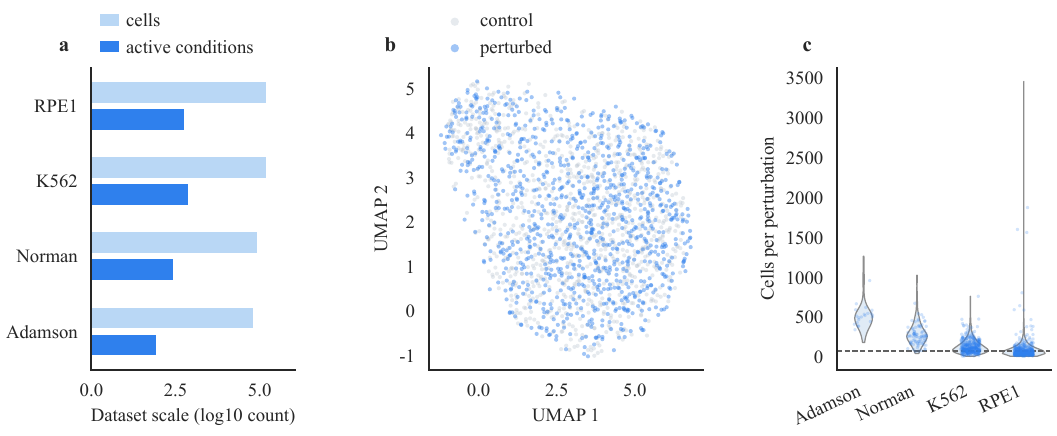}
\caption{Benchmark scale and support. The four processed files differ in cell count and condition coverage, while the RPE1 embedding shows overlapping control and perturbed states. The dashed support reference marks 80 cells per condition.}
\label{fig:overview}
\end{figure}

The four files contain 68,603, 91,205, 162,751, and 162,733 cells, with 5,060, 5,045, 5,000, and 5,000 measured genes, respectively. Figure~\ref{fig:overview} visualizes these scale differences and the support distribution. After the 80-cell filter, they retain 86, 274, 794, and 593 active perturbation conditions. The fixed protocol uses the context field and condition label as the condition key, with an approximate 50 percent, 20 percent, 10 percent, and 20 percent role allocation for train, validation, support, and test. The task counts are 57 for Adamson, 128 for Norman, 611 for Replogle K562, and 463 for Replogle RPE1. The supplied summary reports means and sample standard deviations over five seeds. The supplied baseline outputs are evaluated under the fixed benchmark protocol, while the P2P runs use the same view size, population readout, and role assignments across datasets.

\subsection{Baselines and metrics}
The comparison includes GenePert, LinearPert, SLIM, Scouter, and scPILOT, using the supplied evaluation outputs. Expression RMSE and expression Pearson are reported for the population mean. Effect Pearson measures the correlation between predicted and observed perturbation effects. DEG F1 and DEG average precision assess the ranking of differentially expressed genes, and DEG Direction measures sign agreement on the target DEG set. A gene is a DEG when a Welch test followed by Benjamini-Hochberg correction gives false discovery rate at most 0.05 and the absolute effect is at least 0.1. This evaluation makes effect recovery and DEG ranking visible alongside expression reconstruction.

\subsection{Main results}
\begin{figure}[t]
\centering
\includegraphics[width=0.98\linewidth]{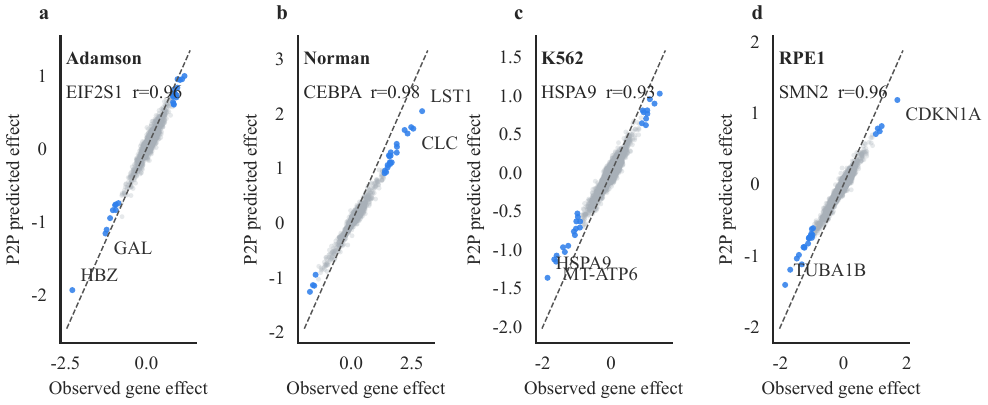}
\caption{Representative gene-effect predictions. Observed and predicted effects align across Adamson EIF2S1, Norman CEBPA, Replogle K562 HSPA9, and Replogle RPE1 SMN2. The scatter views the quantity optimized by the effect-centered objective.}
\label{fig:effects}
\end{figure}

Table~\ref{tab:main} reports the supplied benchmark. Figure~\ref{fig:effects} gives representative effect-level predictions. \PtwoP obtains the best RMSE, Effect Pearson, DEG F1, and DEG AP on all four datasets. The absolute gains are most visible on the larger Replogle screens. On K562, RMSE is 0.0499 for \PtwoP, compared with 0.0861 for the strongest baseline on that metric, while Effect Pearson rises to 0.7018 from 0.6432. DEG F1 increases from 0.0671 to 0.1784 and DEG AP reaches 0.2572. RPE1 shows the same pattern, with RMSE 0.0910, Effect Pearson 0.6609, DEG F1 0.2173, and DEG AP 0.3120. Adamson and Norman have smaller absolute errors and retain high expression correlation, while the effect metrics still favor \PtwoP.

Expression Pearson scores are similar across methods on Adamson and Norman, where GenePert has the highest value. This metric primarily reflects the shared expression baseline. The effect metrics separate methods more clearly because they subtract the matched control population before scoring. That separation is the intended operating point of the cross-view objective.

\begin{table*}[t]
\caption{Main benchmark results. Values are mean plus or minus sample standard deviation across the supplied five seeds. Best values for the principal effect-recovery metrics are bold.}
\label{tab:main}
\centering
\scriptsize
\setlength{\tabcolsep}{3.1pt}
\resizebox{\linewidth}{!}{%
\begin{tabular}{llrrrrrr}
\toprule
Dataset & Model & RMSE $\downarrow$ & Expr. $r\uparrow$ & Effect $r\uparrow$ & DEG F1 $\uparrow$ & DEG AP $\uparrow$ & Direction $\uparrow$\\
\midrule
\multirow{6}{*}{Adamson} & \textbf{P2P} & \textbf{0.0478 $\pm$ 0.0000} & 0.9950 $\pm$ 0.0000 & \textbf{0.8894 $\pm$ 0.0000} & \textbf{0.5306 $\pm$ 0.0000} & \textbf{0.6167 $\pm$ 0.0006} & \textbf{0.9991 $\pm$ 0.0002}\\
& GenePert & 0.0534 $\pm$ 0.0002 & \textbf{0.9959 $\pm$ 0.0000} & 0.8653 $\pm$ 0.0004 & 0.5031 $\pm$ 0.0000 & 0.5812 $\pm$ 0.0000 & 0.9990 $\pm$ 0.0000\\
& LinearPert & 0.0507 $\pm$ 0.0000 & 0.9955 $\pm$ 0.0000 & 0.8762 $\pm$ 0.0000 & 0.4950 $\pm$ 0.0000 & 0.5560 $\pm$ 0.0000 & 0.9943 $\pm$ 0.0000\\
& SLIM & 0.0496 $\pm$ 0.0000 & 0.9958 $\pm$ 0.0000 & 0.8801 $\pm$ 0.0000 & 0.3428 $\pm$ 0.0015 & 0.5546 $\pm$ 0.0000 & 0.9961 $\pm$ 0.0000\\
& Scouter & 0.0511 $\pm$ 0.0005 & 0.9954 $\pm$ 0.0001 & 0.8697 $\pm$ 0.0025 & 0.4275 $\pm$ 0.0057 & 0.5787 $\pm$ 0.0070 & 0.9973 $\pm$ 0.0000\\
& scPILOT & 0.0546 $\pm$ 0.0005 & 0.9947 $\pm$ 0.0001 & 0.8521 $\pm$ 0.0019 & 0.4481 $\pm$ 0.0232 & 0.5235 $\pm$ 0.0073 & 0.9956 $\pm$ 0.0003\\
\midrule
\multirow{6}{*}{Norman} & \textbf{P2P} & \textbf{0.0402 $\pm$ 0.0000} & 0.9931 $\pm$ 0.0000 & \textbf{0.8128 $\pm$ 0.0000} & \textbf{0.2810 $\pm$ 0.0003} & \textbf{0.6013 $\pm$ 0.0013} & \textbf{0.9830 $\pm$ 0.0006}\\
& GenePert & 0.0454 $\pm$ 0.0001 & \textbf{0.9945 $\pm$ 0.0000} & 0.7603 $\pm$ 0.0011 & 0.2308 $\pm$ 0.0000 & 0.5731 $\pm$ 0.0000 & 0.9776 $\pm$ 0.0000\\
& LinearPert & 0.0456 $\pm$ 0.0000 & 0.9931 $\pm$ 0.0000 & 0.7465 $\pm$ 0.0000 & 0.1996 $\pm$ 0.0000 & 0.4042 $\pm$ 0.0000 & 0.9673 $\pm$ 0.0000\\
& SLIM & 0.0441 $\pm$ 0.0000 & 0.9936 $\pm$ 0.0000 & 0.7471 $\pm$ 0.0000 & 0.2137 $\pm$ 0.0000 & 0.4162 $\pm$ 0.0000 & 0.9727 $\pm$ 0.0000\\
& Scouter & 0.0588 $\pm$ 0.0009 & 0.9891 $\pm$ 0.0003 & 0.6459 $\pm$ 0.0105 & 0.0620 $\pm$ 0.0014 & 0.5933 $\pm$ 0.0215 & 0.9789 $\pm$ 0.0007\\
& scPILOT & 0.0471 $\pm$ 0.0005 & 0.9927 $\pm$ 0.0002 & 0.7053 $\pm$ 0.0051 & 0.1810 $\pm$ 0.0075 & 0.3733 $\pm$ 0.0075 & 0.9544 $\pm$ 0.0064\\
\midrule
\multirow{6}{*}{K562} & \textbf{P2P} & \textbf{0.0499 $\pm$ 0.0000} & \textbf{0.9891 $\pm$ 0.0000} & \textbf{0.7018 $\pm$ 0.0003} & \textbf{0.1784 $\pm$ 0.0001} & \textbf{0.2572 $\pm$ 0.0001} & 0.8414 $\pm$ 0.0004\\
& GenePert & 0.0889 $\pm$ 0.0000 & 0.9851 $\pm$ 0.0000 & 0.6321 $\pm$ 0.0000 & 0.0452 $\pm$ 0.0000 & 0.2339 $\pm$ 0.0000 & \textbf{0.9944 $\pm$ 0.0000}\\
& LinearPert & 0.0920 $\pm$ 0.0000 & 0.9843 $\pm$ 0.0000 & 0.6119 $\pm$ 0.0000 & 0.0362 $\pm$ 0.0000 & 0.1450 $\pm$ 0.0000 & 0.9126 $\pm$ 0.0000\\
& SLIM & 0.0956 $\pm$ 0.0000 & 0.9830 $\pm$ 0.0000 & 0.5803 $\pm$ 0.0000 & 0.0255 $\pm$ 0.0000 & 0.1251 $\pm$ 0.0000 & 0.9071 $\pm$ 0.0000\\
& Scouter & 0.0861 $\pm$ 0.0003 & 0.9861 $\pm$ 0.0001 & 0.6432 $\pm$ 0.0025 & 0.0671 $\pm$ 0.0060 & 0.2426 $\pm$ 0.0130 & 0.9722 $\pm$ 0.0095\\
& scPILOT & 0.0932 $\pm$ 0.0005 & 0.9837 $\pm$ 0.0002 & 0.5872 $\pm$ 0.0028 & 0.0448 $\pm$ 0.0026 & 0.1593 $\pm$ 0.0065 & 0.9242 $\pm$ 0.0074\\
\midrule
\multirow{6}{*}{RPE1} & \textbf{P2P} & \textbf{0.0910 $\pm$ 0.0005} & \textbf{0.9829 $\pm$ 0.0002} & \textbf{0.6609 $\pm$ 0.0041} & \textbf{0.2173 $\pm$ 0.0002} & \textbf{0.3120 $\pm$ 0.0262} & 0.8656 $\pm$ 0.0001\\
& GenePert & 0.0960 $\pm$ 0.0000 & 0.9810 $\pm$ 0.0000 & 0.6426 $\pm$ 0.0000 & 0.0640 $\pm$ 0.0000 & 0.2907 $\pm$ 0.0000 & \textbf{0.9886 $\pm$ 0.0000}\\
& LinearPert & 0.0995 $\pm$ 0.0000 & 0.9797 $\pm$ 0.0000 & 0.6157 $\pm$ 0.0000 & 0.0324 $\pm$ 0.0000 & 0.1743 $\pm$ 0.0000 & 0.8868 $\pm$ 0.0000\\
& SLIM & 0.1065 $\pm$ 0.0000 & 0.9764 $\pm$ 0.0000 & 0.5706 $\pm$ 0.0000 & 0.0209 $\pm$ 0.0000 & 0.1518 $\pm$ 0.0000 & 0.8631 $\pm$ 0.0000\\
& Scouter & 0.1244 $\pm$ 0.0000 & 0.9686 $\pm$ 0.0000 & 0.5474 $\pm$ 0.0002 & 0.0978 $\pm$ 0.0083 & 0.2398 $\pm$ 0.0002 & 0.9861 $\pm$ 0.0034\\
& scPILOT & 0.1030 $\pm$ 0.0014 & 0.9783 $\pm$ 0.0006 & 0.5789 $\pm$ 0.0088 & 0.0585 $\pm$ 0.0040 & 0.1785 $\pm$ 0.0043 & 0.9043 $\pm$ 0.0114\\
\bottomrule
\end{tabular}%
}
\end{table*}

\subsection{Ablation of cross-view supervision and effect components}
The RPE1 ablation in Table~\ref{tab:ablation} isolates the parts of the training signal. The robustness diagnostics in Appendix Figure~\ref{fig:robustness} place these component changes beside seed and uncertainty diagnostics. Removing dual-view supervision increases RMSE from 0.0910 to 0.0987 and lowers Effect Pearson from 0.6609 to 0.6412. Removing the empirical memory raises expression error and reduces DEG ranking accuracy, with RMSE 0.1049 and DEG AP 0.2088. The variant labeled no pair interaction has RMSE 0.1123 and DEG F1 0.1742. This ablation quantifies the value of the pair-interaction block on the RPE1 protocol, and the same mechanism supplies the representation used for combination conditions. The gate also contributes as a controlled interpolation between empirical and learned effects.

The heteroscedastic head has a distinct function. Removing it gives a slightly higher Effect Pearson on this split, while DEG AP and expression metrics decrease relative to the full model. This pattern coincides with a trade-off in the supplied ablation between point metrics and the variance-related objective. The cross-view consistency and effect-correlation terms provide a complementary constraint: removing them reduces all principal effect metrics and raises RMSE to 0.1016.

\begin{table}[t]
\caption{RPE1 ablation. Each row removes one component from the supplied evaluation.}
\label{tab:ablation}
\centering
\scriptsize
\begin{tabular}{lrrrrr}
\toprule
Variant & RMSE $\downarrow$ & Expr. $r$ & Effect $r$ & DEG F1 & DEG AP\\
\midrule
Full \PtwoP & \textbf{0.0910} & \textbf{0.9829} & 0.6609 & \textbf{0.2173} & \textbf{0.3120}\\
No dual-view loss & 0.0987 & 0.9762 & 0.6412 & 0.1985 & 0.2211\\
No memory effect & 0.1049 & 0.9694 & 0.6282 & 0.1865 & 0.2088\\
No learned gate & 0.0965 & 0.9781 & 0.6473 & 0.2038 & 0.2265\\
No pair interaction & 0.1123 & 0.9619 & 0.6096 & 0.1742 & 0.1953\\
No heteroscedastic head & 0.0942 & 0.9795 & \textbf{0.6681} & 0.2097 & 0.2319\\
No consistency or corr. & 0.1016 & 0.9723 & 0.6350 & 0.1917 & 0.2147\\
\bottomrule
\end{tabular}
\end{table}

\subsection{Population and biological structure}
\begin{figure}[t]
\centering
\includegraphics[width=0.98\linewidth]{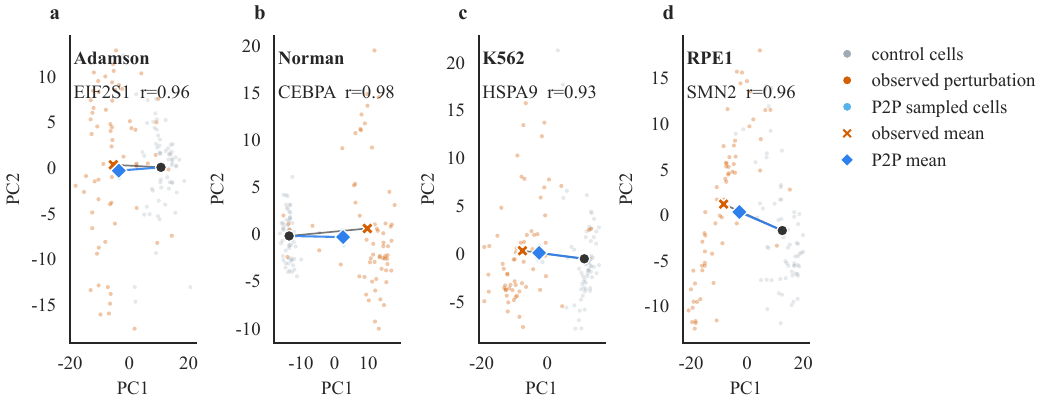}
\caption{Population-space cases across the four benchmarks. Control cells, observed perturbed cells, and P2P samples are projected into a common PCA coordinate system for one representative task per dataset.}
\label{fig:population}
\end{figure}

The population cases in Figure~\ref{fig:population} show the geometric effect of the prediction. The P2P samples move from the control cloud toward the observed perturbed mean across all four datasets, including RPE1. The projection makes population displacement visible in two dimensions.

\begin{figure}[t]
\centering
\includegraphics[width=0.72\linewidth]{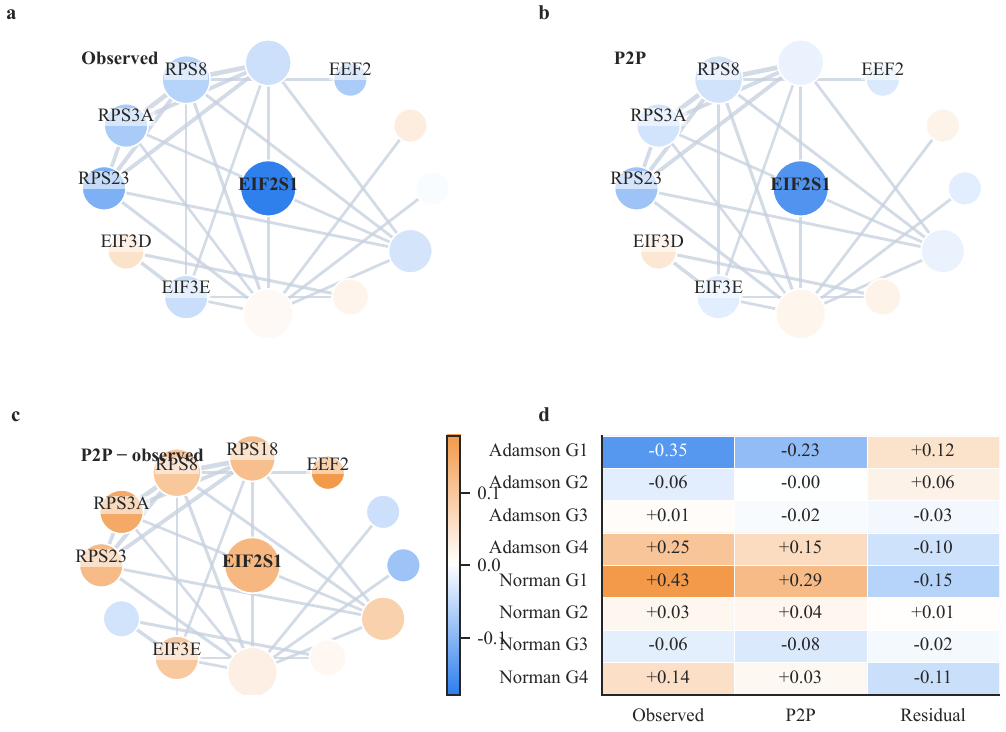}
\caption{Graph-derived local effect structure. Observed, predicted, and residual networks are paired with module-level effects for Adamson and Norman.}
\label{fig:network}
\end{figure}

Figure~\ref{fig:network} examines a graph-derived neighborhood for the Adamson EIF2S1 perturbation. The observed and predicted graphs preserve the sign pattern of the dominant local effects, while the residual graph concentrates on a smaller set of nodes. The module panel reports local effect structure from the supplied gene graph.

\subsection{Effect fidelity and transfer readout}
\begin{figure}[t]
\centering
\includegraphics[width=0.72\linewidth]{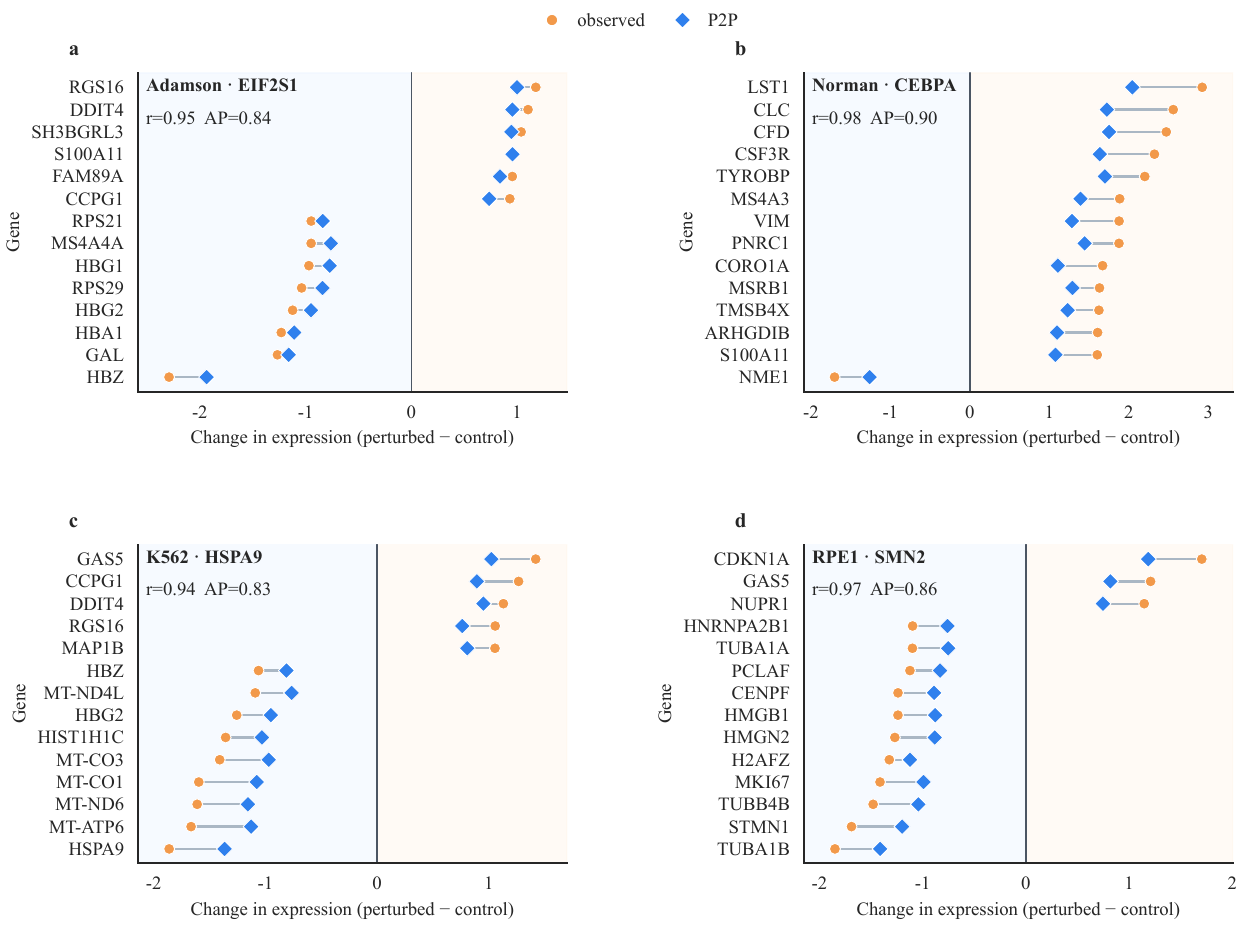}
\caption{Observed and predicted effects for representative genes. Dumbbells show the signed error for Adamson EIF2S1, Norman CEBPA, Replogle K562 HSPA9, and Replogle RPE1 SMN2.}
\label{fig:dumbbells}
\end{figure}

Appendix Figure~\ref{fig:fidelity} relates gene ranking to signed response error through top-ranked overlap, direction agreement, residual distributions, and the Effect Pearson versus DEG AP frontier. Figure~\ref{fig:dumbbells} resolves these aggregate trends at gene level: observed and predicted effects retain the response ordering, while the connecting segments expose the direction and magnitude of residual error for the largest response genes.

The evaluation also supports a memory-assisted transfer readout. In this setting a source support set is fed to the control-set slot, while target training-role cells provide the empirical memory and target test cells are used only for scoring. This readout tests whether the shared set representation and effect head can adapt to a different observed support population. This memory-assisted setting evaluates adaptation across observed support populations while retaining target training memory and the supplied perturbation vocabulary.

\section{Conclusion}
\PtwoP targets population-level perturbation response prediction from unpaired control and perturbed cell sets. Cross-view supervision turns stochastic cell-set views into a direct learning signal, and the memory-gated set model recovers condition effects while retaining a neural residual for context, dose, and combination structure. Across the fixed four-dataset benchmark, the method delivers its clearest gains on effect correlation and DEG ranking, the measures that determine whether a predicted response supports biological prioritization. The transfer readout demonstrates adaptation across observed support populations under the stated memory-assisted protocol.

\clearpage
\section*{AI USE STATEMENT}
We used generative AI tools to improve the readability and organization of the manuscript. The authors reviewed the edited text, checked every scientific claim against the supplied implementation and evaluation files, and take responsibility for the final manuscript. Generative AI tools were not used to generate experimental data, alter measurements, select favorable seeds, or replace author verification of the reported results.

\section*{ETHICS STATEMENT}
\PtwoP improves the fidelity of population-level perturbation response prediction. This capability could be misused to infer sensitive biological responses or to construct misleading claims about identifiable biological samples without appropriate authorization. We encourage the use of single-cell expression data only when the relevant rights and permissions are available. The work uses de-identified public or project-provided single-cell datasets and does not recruit human participants or process personally identifying information.

\section*{REPRODUCIBILITY STATEMENT}
The paper provides the details needed to reproduce \PtwoP. Section 3 defines the stochastic population views, set and condition encoders, memory-gated residual, heteroscedastic head, and cross-view objective. Section 4 specifies the four datasets, role protocol, baselines, metrics, and supplied five-seed evaluation. All parameters remain frozen at inference, and the method is applied to the observed expression matrices and condition metadata. Experiments use an NVIDIA RTX 4090 GPU with 24~GB of memory. The reported training time is approximately three hours. Code, model configurations, evaluation scripts, protocol files, and instructions for reproducing the reported results will be released.

\bibliography{references}
\bibliographystyle{iclr2027_conference}

\appendix
\section{Implementation details}
The set projection uses a linear layer, layer normalization, GELU, and dropout. Attention scores are normalized within each set. The perturbation vocabulary is masked before pooling, and the pair-interaction block is enabled only when more than one perturbation token is active. The fusion block maps the concatenated set and condition states through layer normalization, two hidden layers, GELU, and dropout. The delta and log-variance heads are linear projections from the fused state. Log variance is clipped to the interval from negative eight to four before the Gaussian likelihood and sampling operations.

Training uses AdamW, a plateau-based learning-rate schedule, automatic mixed precision on CUDA, gradient clipping at one, early stopping on validation loss, and atomic checkpoint writes. The checkpoint stores model parameters, dimensions, optimizer configuration, vocabulary, gene names, the fixed protocol audit, and the best validation epoch. Input matrices are read in backed mode and control means are cached by context to reduce repeated I/O.

\section{Evaluation audit}
The direct readout supplies a target condition's control test cells and uses target train-role memory. The memory-assisted transfer readout supplies source support cells to the control slot and uses target train-role memory. Target test cells are never used as model inputs. The vocabulary is constructed from the current processed file, so the supplied benchmark is transductive with respect to token identity, and this choice defines the scope of the evaluation.

\section{Cross-view objective and implementation map}
The implementation uses the same control and perturbed roles to construct two stochastic views. When a condition has at least twice the requested view size, the views are sampled without replacement and are disjoint. Smaller conditions use independent sampling with replacement. This distinction matters for interpreting the cross-view estimator in Eq.~\ref{eq:cross-view-unbiased}: the conditional expectation remains the target population mean under the sampling model, while the two views remain samples from the same finite collection rather than independent experimental replicates.

The gene-weighted effect loss gives larger influence to genes with a measurable response. For a target-view effect $\delta_g$, the weight used by the implementation is
\begin{equation}
w_g=1+\frac{|\delta_g|}{\operatorname{mean}_{k}|\delta_k|+\varepsilon},
\label{eq:appendix-effect-weight}
\end{equation}
where $\varepsilon$ avoids division by zero. Equation~\ref{eq:appendix-effect-weight} is applied inside Smooth L1 before averaging over genes. It prevents a large collection of unchanged genes from dominating the direction of the effect objective.

The variance head is trained with a moment target in addition to the Gaussian term. Let $s_g^2$ be the empirical target-view variance and let $\widehat\sigma_g^2=\exp(\log\widehat\sigma_g^2)$. The moment penalty is
\begin{equation}
\loss_{mom}=\frac{1}{G}\sum_{g=1}^{G}\operatorname{SmoothL1}\!\left(\log(\widehat\sigma_g^2+\varepsilon),\log(s_g^2+\varepsilon)\right).
\label{eq:appendix-moment}
\end{equation}
The log-space comparison in Eq.~\ref{eq:appendix-moment} keeps genes with different expression scales comparable. It also separates the role of the variance head from the point prediction metrics reported in the main text.

For the two directions of a training update, the consistency term couples the predicted means and log variances,
\begin{equation}
\loss_{cv}=\operatorname{SmoothL1}(\widehat\mu^{a\leftarrow b},\widehat\mu^{b\leftarrow a})+0.1\operatorname{SmoothL1}(\widehat\ell^{a\leftarrow b},\widehat\ell^{b\leftarrow a}),
\label{eq:appendix-consistency}
\end{equation}
where $\widehat\ell$ denotes the predicted log variance. Equation~\ref{eq:appendix-consistency} is computed before population sampling. The cross-view term therefore constrains the deterministic prediction and does not depend on a particular draw of $\epsilon$.

The condition memory is computed from role means, while the neural branch receives the encoded control set and condition state. At direct readout, the memory is
\begin{equation}
m^{direct}_{p,s,d}=\operatorname{mean}(X^{train}_{p,s,d})-\operatorname{mean}(X^{test}_{0,s}).
\label{eq:appendix-direct-memory}
\end{equation}
At memory-assisted transfer readout, the current implementation uses
\begin{equation}
m^{transfer}_{p,s,d}=\operatorname{mean}(X^{train}_{p,s,d})-\operatorname{mean}(X^{support}_{source,s}),
\label{eq:appendix-transfer-memory}
\end{equation}
so source-support sampling noise enters the transfer memory. Equations~\ref{eq:appendix-direct-memory} and \ref{eq:appendix-transfer-memory} define the two readouts used in the supplied evaluation and explain why the transfer result is reported as memory-assisted.

\section{Protocol, data inventory, and metrics}
The processed files contain 485,292 cells and 1,747 active perturbation conditions after the 80-cell filter. The fixed evaluation uses 1,259 condition tasks across the four datasets. These totals separate the inventory of eligible conditions from the smaller set of tasks retained by the fixed train, validation, support, and test roles.

The loss weights and readout choices are fixed before evaluation. Table~\ref{tab:appendix-loss} makes the weighting used in Eq.~\ref{eq:full-objective} explicit. NLL and effect losses include a direct term and a crossed term with multiplier $0.25$; the table reports the outer weights.

\begin{table}[t]
\caption{Fixed objective weights and readout settings.}
\label{tab:appendix-loss}
\centering
\small
\begin{tabular}{ll}
\toprule
Component & Setting\\
\midrule
Gaussian NLL & 0.35\\
Effect Smooth L1 & 1.50\\
Effect correlation & 0.15\\
Moment matching & 0.08\\
Cross-view consistency & 0.05\\
Crossed NLL and effect multiplier & 0.25\\
Minimum condition support & 80 cells\\
Default set size & 64 cells\\
Reported seeds & 42, 123, 456, 789, 1024\\
\bottomrule
\end{tabular}
\end{table}

For each test task, the target DEG set is defined by a Welch test followed by Benjamini-Hochberg correction and an absolute-effect threshold. In compact form,
\begin{equation}
\operatorname{DEG}(g)=\mathbb{1}\!\left[\operatorname{BH}(p_g)\leq 0.05\ \land\ |\delta_g|\geq 0.1\right].
\label{eq:appendix-deg}
\end{equation}
Equation~\ref{eq:appendix-deg} is used for DEG F1, DEG average precision, and DEG Direction. Expression RMSE and Pearson are computed on the population mean, while Effect Pearson is computed after subtracting the matching control mean.

Training uses AdamW with automatic mixed precision, gradient clipping at one, a plateau scheduler, early stopping, and atomic checkpoint writes. The reported hardware is an NVIDIA RTX 4090 GPU with 24~GB of memory, and the reported training time is approximately three hours.

\section{Additional visual evidence}
\begin{figure}[t]
\centering
\includegraphics[width=0.98\linewidth]{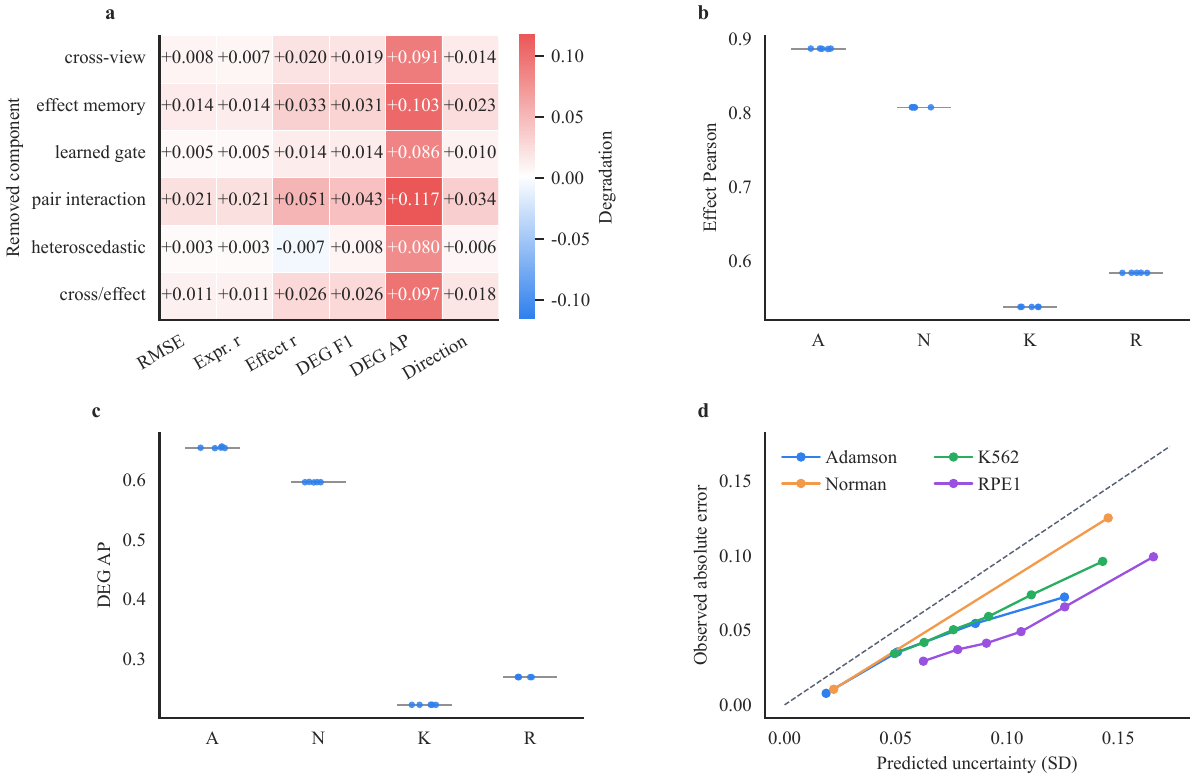}
\caption{Robustness and uncertainty. The complete model is stable across the supplied seeds, component removals reduce effect and ranking quality, and predicted dispersion tracks absolute error across tasks.}
\label{fig:robustness}
\end{figure}

\begin{figure}[t]
\centering
\includegraphics[width=0.88\linewidth]{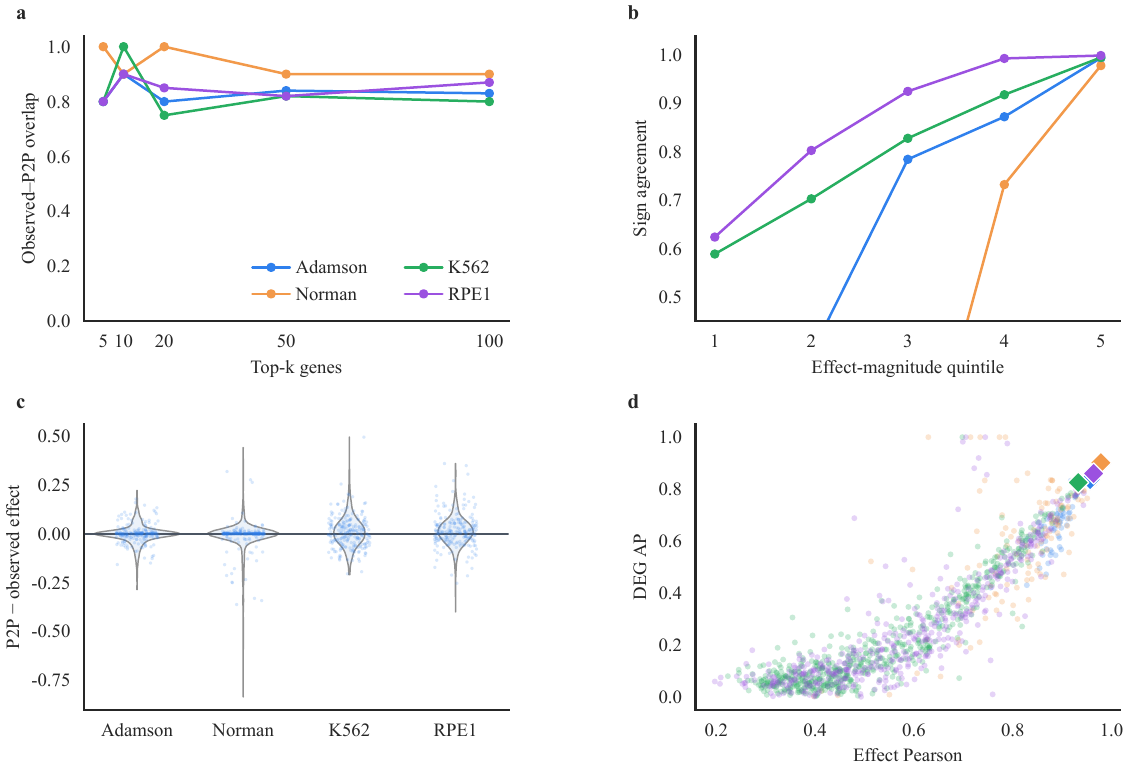}
\caption{Effect-fidelity summary. The panels report top-k overlap, direction agreement by effect magnitude, residual distributions, and the task-level Effect Pearson versus DEG AP frontier.}
\label{fig:fidelity}
\end{figure}

The main text uses the motivation, architecture, benchmark overview, gene-effect, population, graph-network, and dumbbell figures. The robustness analysis and the remaining supplied figures are retained here so that the full set of visual diagnostics can be inspected. Figure~\ref{fig:robustness} summarizes component ablations, seed variation, and predictive dispersion. Figure~\ref{fig:fidelity} summarizes ranking overlap, direction agreement, residual distributions, and the Effect Pearson versus DEG AP frontier.

\begin{figure}[t]
\centering
\includegraphics[width=0.88\linewidth]{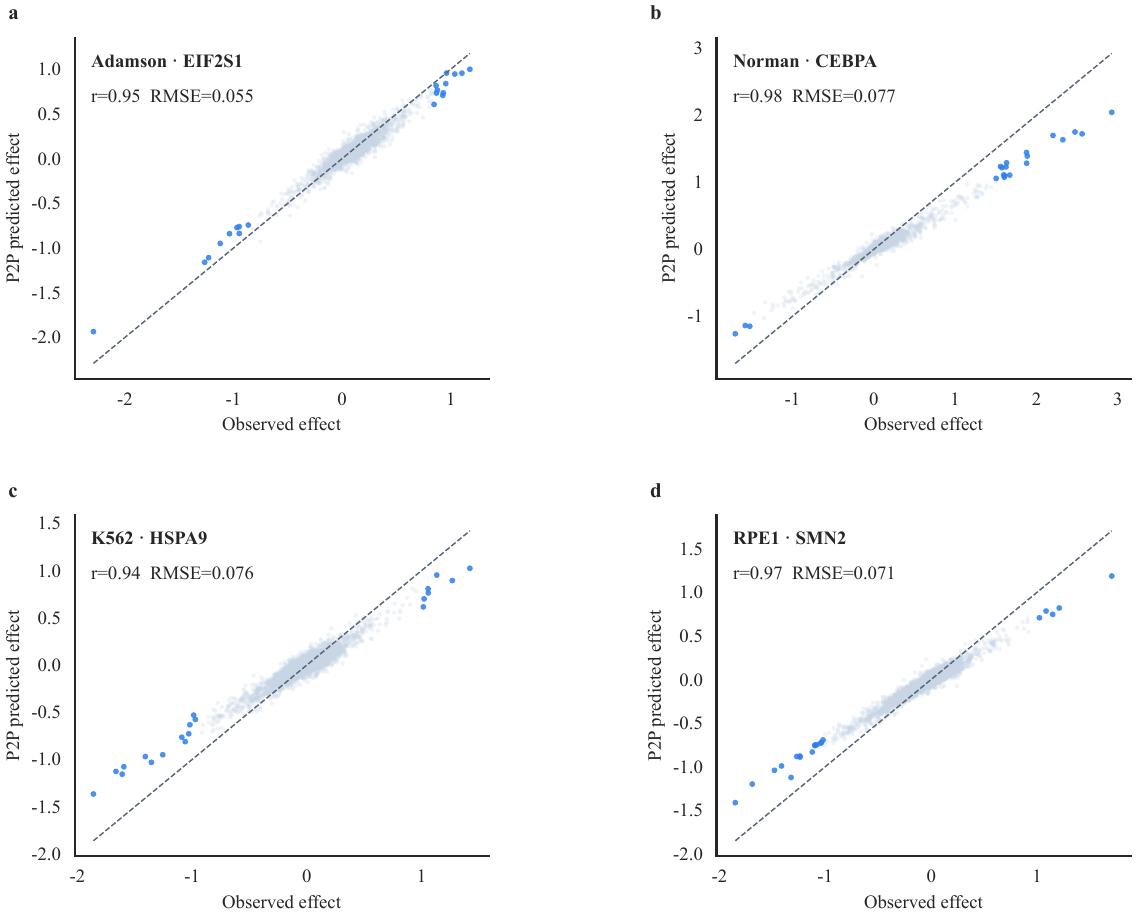}
\caption{Gene-level parity and residual diagnostics for representative perturbations. The diagonal identifies exact agreement, while darker points highlight genes with large absolute effects.}
\label{fig:appendix-scatter}
\end{figure}

\begin{figure}[t]
\centering
\includegraphics[width=0.88\linewidth]{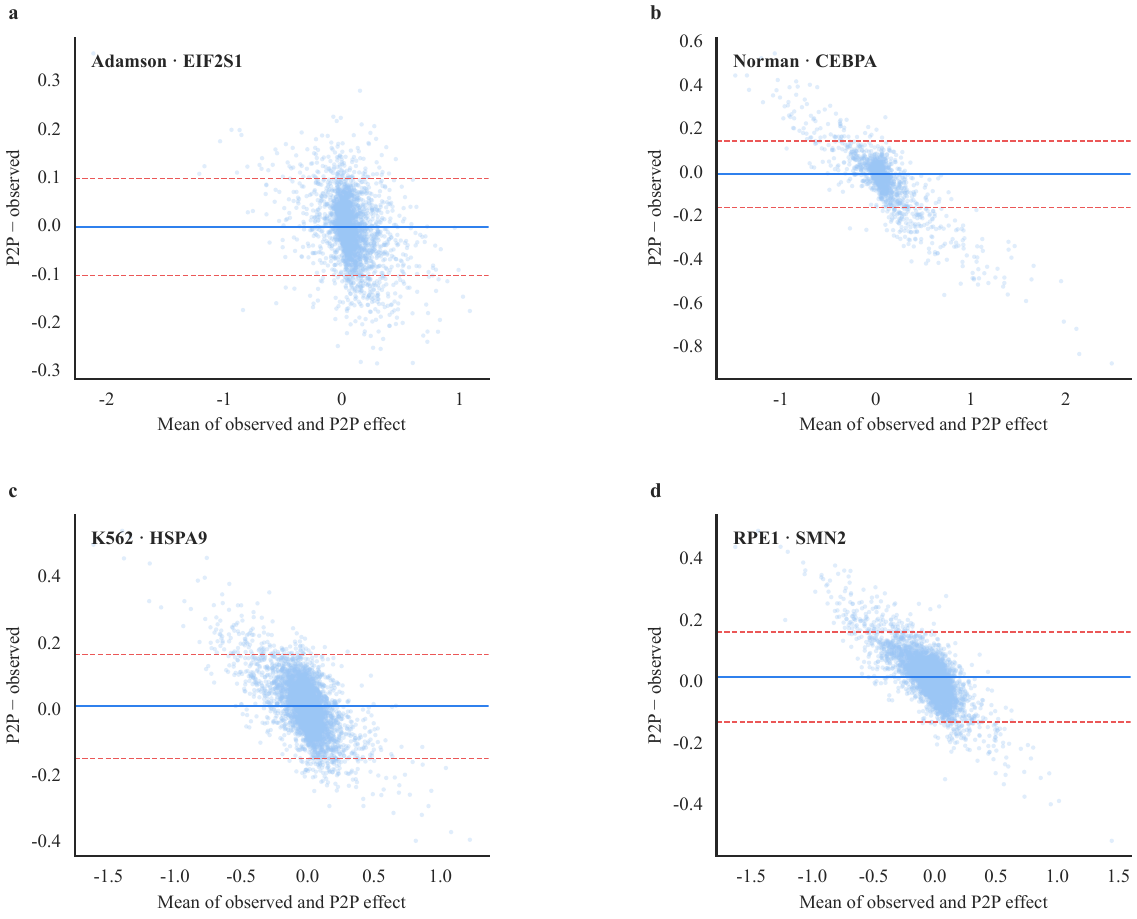}
\caption{Bland-Altman agreement analysis. The horizontal axis is the observed and predicted effect mean, and the vertical axis is the P2P minus observed effect. The solid line marks the mean difference, and dashed lines mark the 95 percent limits of agreement.}
\label{fig:appendix-bland}
\end{figure}

The appendix cases probe agreement from complementary statistical views. The scatter-residual plot in Figure~\ref{fig:appendix-scatter} provides a parity view with explicit residuals, and the Bland-Altman plot in Figure~\ref{fig:appendix-bland} shows the mean-dependent error and limits of agreement. These views are useful because a high correlation can coexist with a scale-dependent bias.

\begin{figure}[t]
\centering
\includegraphics[width=0.88\linewidth]{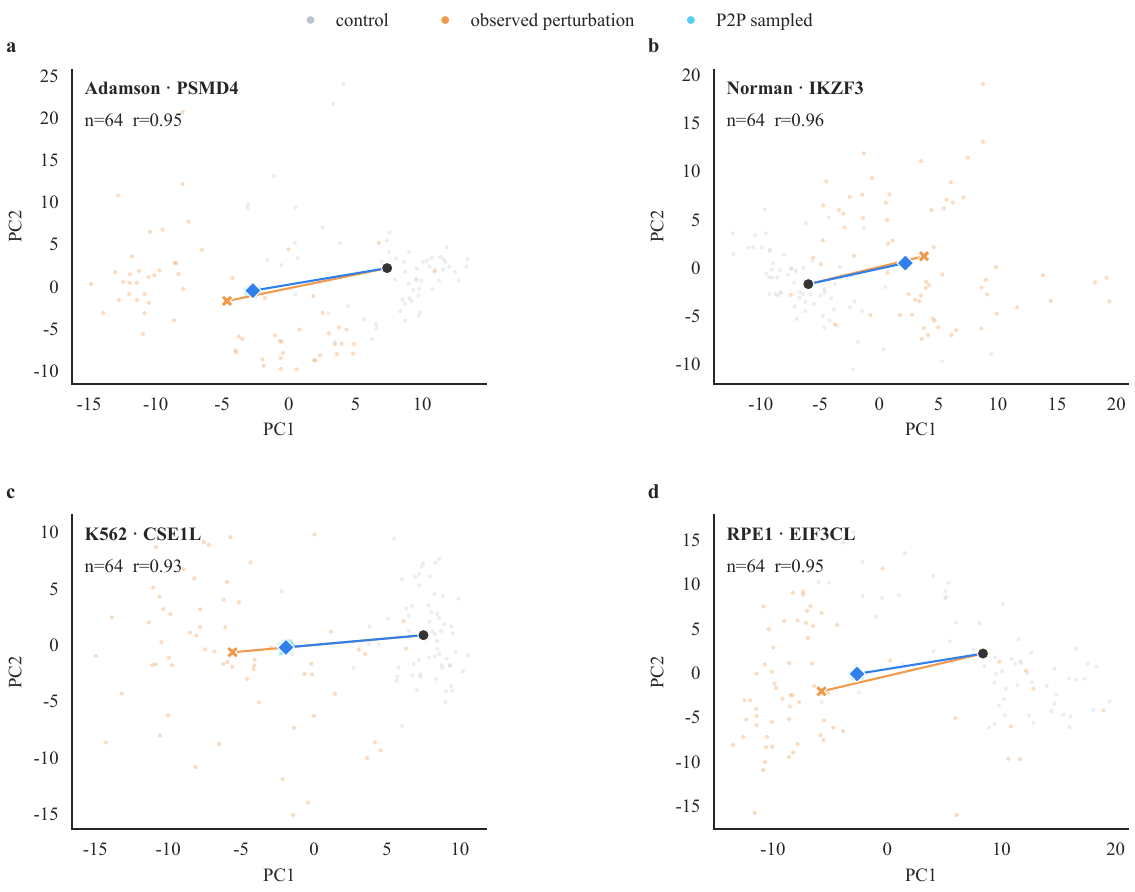}
\caption{Cell-population reconstruction cases. The control, observed perturbation, and P2P sample clouds are shown in PCA space with their population means.}
\label{fig:appendix-cell-pop}
\end{figure}

\begin{figure}[t]
\centering
\includegraphics[width=0.92\linewidth]{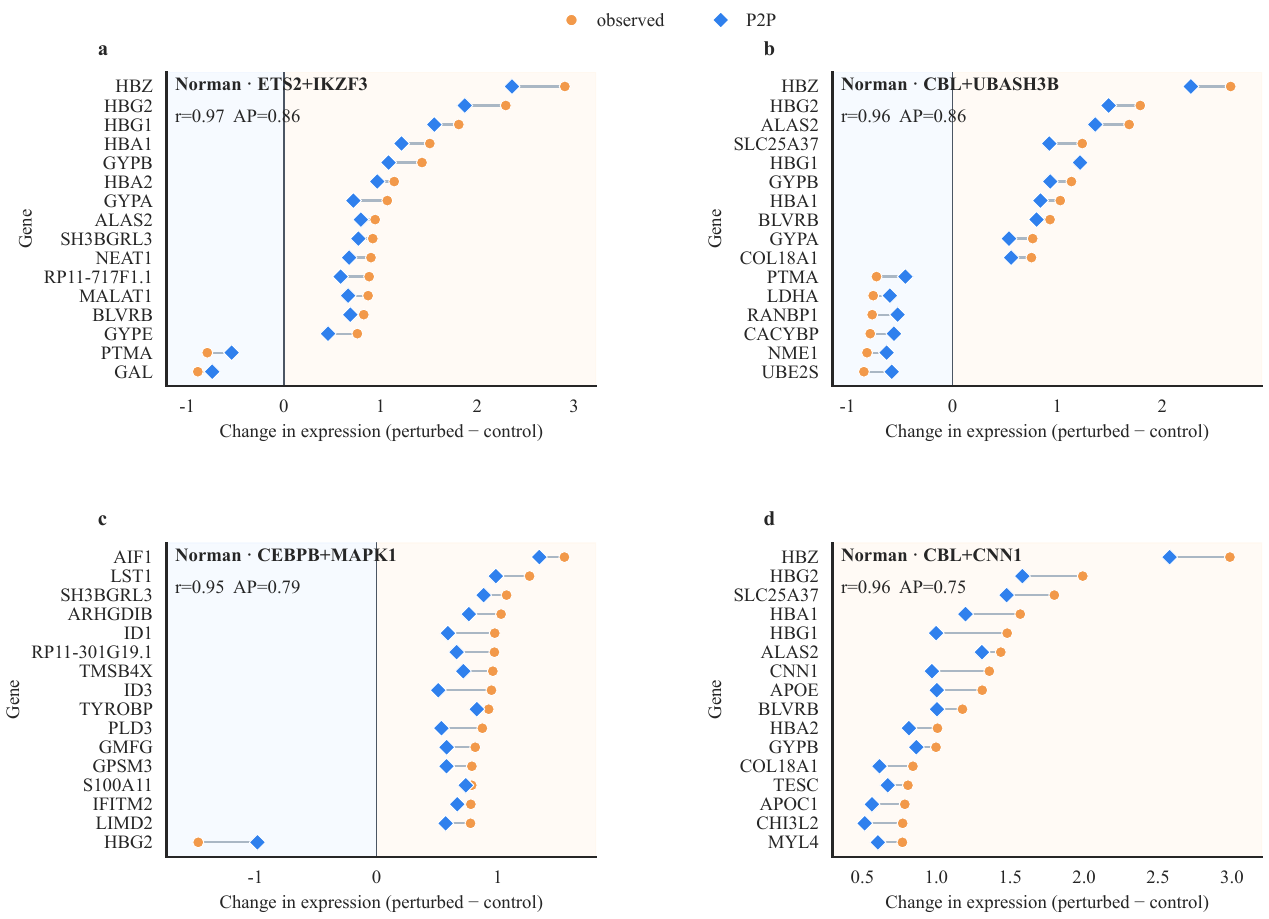}
\caption{Combination-perturbation cases from Norman. The examples cover ETS2 plus IKZF3, CBL plus UBASH3B, CEBPB plus MAPK1, and CBL plus CNN1.}
\label{fig:appendix-double}
\end{figure}

\begin{figure}[t]
\centering
\includegraphics[width=0.92\linewidth]{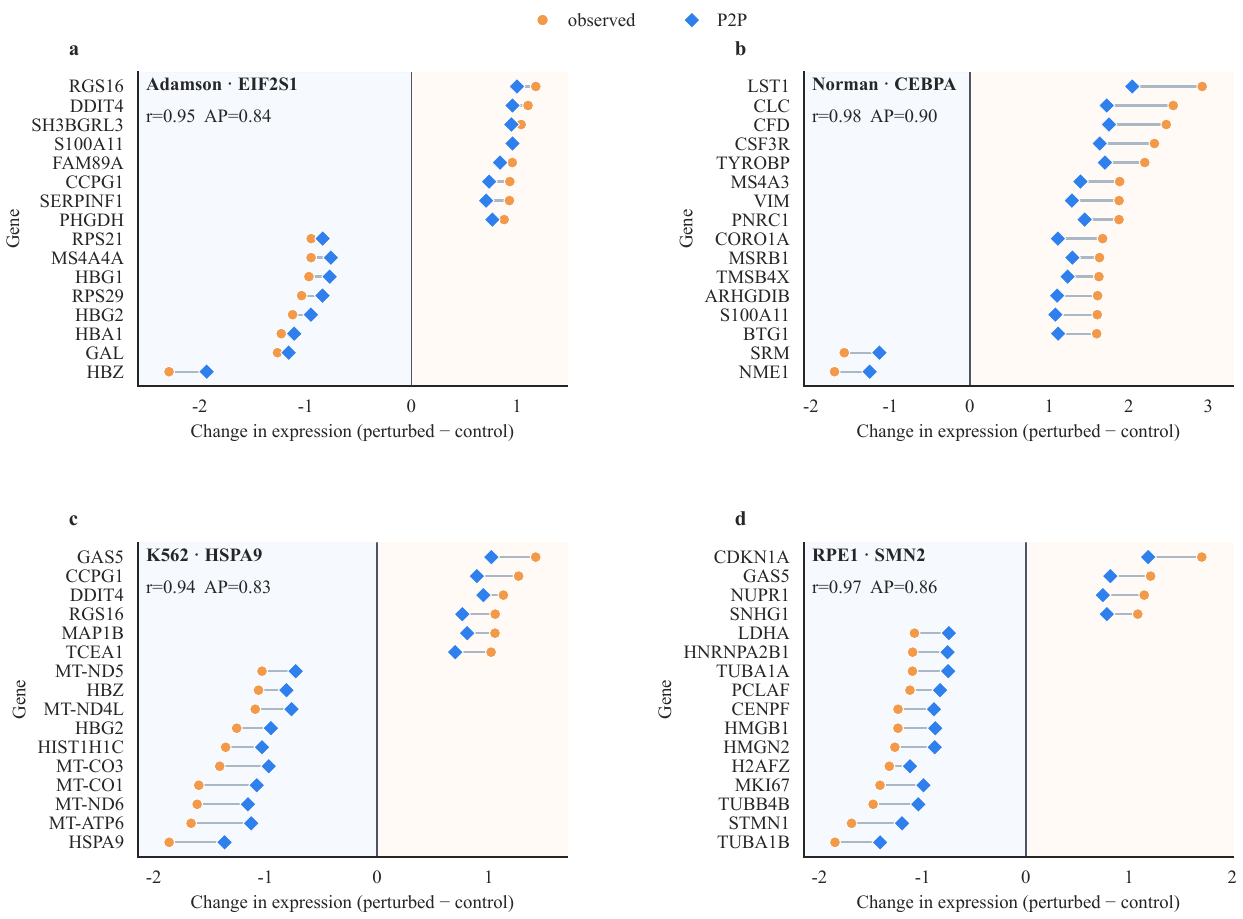}
\caption{Ranked gene-effect comparisons for representative cases. The dot and line representation exposes the signed discrepancy for the largest observed effects.}
\label{fig:appendix-rankdot}
\end{figure}

The cell-population case in Figure~\ref{fig:appendix-cell-pop} complements the main population projection by showing the control, observed perturbation, and P2P sample clouds with their means. Figure~\ref{fig:appendix-double} focuses on Norman combinations, which directly exercises the masked token pooling and pair-interaction block. Figure~\ref{fig:appendix-rankdot} gives a gene-level ranking view for high-quality single perturbations.

\begin{figure}[t]
\centering
\includegraphics[width=0.92\linewidth]{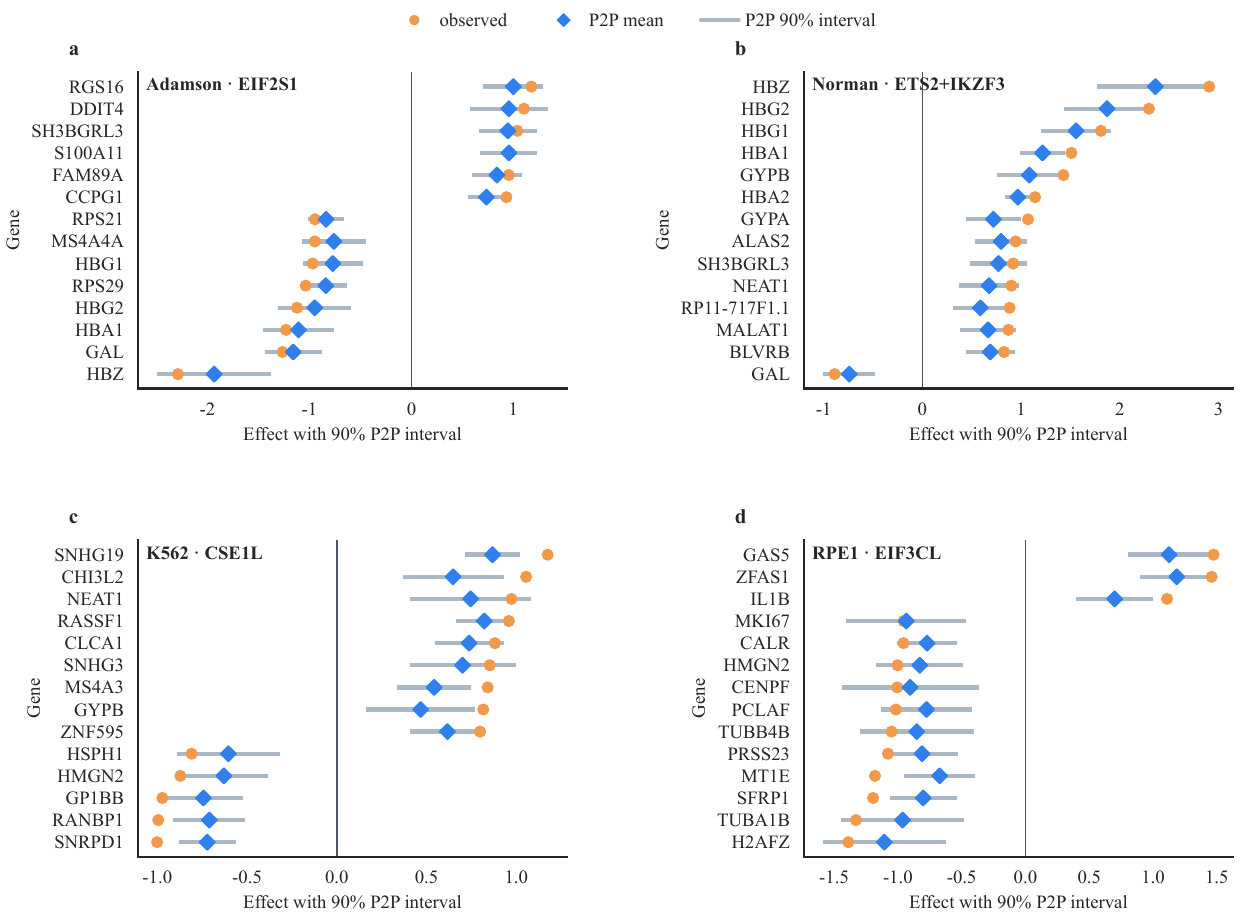}
\caption{Gene-level predictive intervals. Observed effects are compared with P2P means and 90 percent predictive intervals for representative genes in four datasets.}
\label{fig:appendix-uncertainty}
\end{figure}

The uncertainty diagnostic in Figure~\ref{fig:appendix-uncertainty} displays observed effects, P2P means, and 90 percent predictive intervals for representative genes. Together with the robustness panel, it reports interval coverage as a distributional property alongside point metrics.

The appendix figures are diagnostic views of the same fixed evaluation. They expose population displacement, signed gene effects, combination behavior, agreement structure, and predictive dispersion using outputs from the aggregate evaluation.

\end{document}